\documentstyle[times,pramana,epsf,floats]{ias}

\def\gsim{\mathrel{
\rlap{\raise 0.511ex \hbox{$>$}}{\lower 0.511ex
\hbox{$\sim$}}}}
\def\lsim{\mathrel{
\rlap{\raise 0.511ex \hbox{$<$}}{\lower 0.511ex
\hbox{$\sim$}}}}

\begin{document}
\mark{{Gravitational Wave Symphony}{B.S. Sathyaprakash}}

\title{The gravitational wave symphony of the Universe}
\author{B.S. Sathyaprakash}
\address{Department of Physics and Astronomy, Cardiff University, 
Cardiff, CF2 3YB, Great Britain}
\begin {abstract}
{
The new millennium will see the upcoming of several ground-based 
interferometric gravitational wave antennas. Within the next
decade a space-based antenna may also begin to observe the
distant Universe. These gravitational wave detectors will
together operate as a network taking data continuously for several
years, watching the transient and continuous phenomena occurring in 
the deep cores of astronomical objects and dense environs of the
early Universe where gravity was extremely strong and highly non-linear.
The network will listen to the waves from rapidly
spinning  non-axisymmetric neutron stars, normal modes of black holes,
binary black hole inspiral and merger, phase transitions in the
early Universe, quantum fluctuations resulting in a characteristic
background in the early Universe. The gravitational wave antennas will open
a new window to observe the dark Universe unreachable via other channels of
astronomical observations. 
}
\end {abstract}
\maketitle

\section{Introduction}

The next few years will witness the opening of the gravitational window 
for observing the Universe using  a world-wide network of cryogenic 
resonant bar and kilometer baseline laser interferometric gravitational
wave (GW) detectors.  These detectors will take data continuously and will 
initially be able to observe inspiralling black holes at a distance 
of 100 Mpc and will enhance their sensitivity by
an order of magnitude, and the volume of search by 3 orders of
magnitude, in about five years after they are built, thus increasing the
number of potential events a thousand-fold. It is this second
phase of operation that will be more interesting from the astrophysical
point of view bringing us a new insight into phenomena poorly understood
or revealing sources never observed at all. The initial phase, however, 
may well see the most luminous sources of radiation, such as 
transient signals arising as a result of the inspiral and merger 
of black hole binaries and from stellar collapse and supernova explosion.
The initial detections, though not expected to be frequent, are 
important from the fundamental physics point of view and will 
enable us to directly test predictions of general relativity that 
have so far eluded all efforts.

We will begin by discussing the nature of GW and
their interaction with matter and how that helps in building GW detectors.  We will then go on to briefly look at the sensitivities
of the upcoming GW antennas and the planned space mission. This will
be followed by a detailed discussion of some of the promising sources of 
GR.  For recent reviews on GW
see Schutz \cite{Schutz 99}, Thorne \cite{Thorne 95}, Flanagan
\cite{Flanagan} and Grishchuk et al. \cite{grishchuk et al}.
Lasota and Marck \cite{lasota.marck} and Bhawal and Iyer 
\cite{bhawal and iyer} contain many articles on GW sources and their detection.

\section {Gravitational Waves: An Overview}
\label{sec:overview}

Gravitational waves result from coherent accelerated motion
of massive objects. Any system with a time varying quadrupole moment
would emit GW.  They interact with particles in their path by
causing an oscillating tidal distortion in proper distances between
the particles. They
are characterized by a dimensionless {\it amplitude} $h.$ The amplitude
$h$ measures the tidal strain $\delta \ell/\ell = h/2$ in the 
proper distance between two free particles separated initially 
by a distance $\ell.$ Just as light and radio waves, 
GW too carry energy and momentum from 
their sources.  Unlike radio waves, however, there is no dipole 
radiation in Einstein's gravity.  The dominant channel for emission is 
quadrupolar.  Since the force of gravity tends to make astronomical 
objects spherical, or at most axisymmetric, GW emission 
is an inefficient process, unless the symmetry is broken by non-axisymmetric 
rotation, presence of magnetic fields, internal stresses due to 
inhomogeneities, an inherent asymmetry in the system, as in binary 
star systems,  etc.  
 
Sometimes astronomical sources of gravitational radiation (GR) can be
very luminous; a supernova explosion at even as great a distance 
as the Virgo super-cluster of galaxies can be as bright as the Moon, 
though only for a short time.  Even so, gravitation being the weakest 
of all known interactions, the waves associated with it are not 
easily detectable. 
Thus, though GW are produced by almost all bodies in motion,
in a majority of the cases the signal strengths will be far 
too small to be sensed with
the current detectors or those that are being built. For example, our
only evidence for the emission of GW comes from the
observation of the decay in the orbital period of the Hulse-Taylor binary 
caused by radiation back-reaction \cite{psr1913}.
Yet the wave amplitude (and the GW frequency) is far 
too small to be sensed by any antenna -- existing, 
under construction or planned.  Only a catastrophic 
astronomical event can lead to a wave amplitude high 
enough to be detectable.

In general relativity the effects of both
the gravitational interaction and GR are described by 
the tidal distortion they produce in the distance between
pairs of free test masses; the Newtonian
inverse-square force field simply determines the background geometry in 
which free particles move along geodesics. The Coulomb tidal force 
due to a gravitating system falls off as $r^{-3}$, whereas the tidal force 
due to GW that it emits falls off as $r^{-1}.$ Thus, 
the gravitational interaction is the dominant effect close to a 
gravitating body and the effect of the waves is pretty much negligible. 
However, far away from the body the effect of the waves are more easily 
sensed than the Coulomb interaction.

There is a fundamental difference in the way GW sources are observed as opposed to electro-magnetic wave sources:
Optical, infrared, radio and
other telescopes are essentially sensitive to the intensity of radiation.
They build up the signal-to-noise ratio (SNR) by incoherent superposition of
waves emitted by a large number of microscopic sources. One, therefore,
only obtains a gross picture of the region emitting the radiation.
The intensity falls off with distance as $r^{-2}$ so that the number of sources
accessible to a telescope that is sensitive to sources with a limiting intensity
$I_0,$ increases as $I_0^{-3/2}$, as the limiting intensity decreases.
The frequency of electro-magnetic radiation is rather large, 
$10^{8}$--$10^{18}$~Hz and one normally deals with a narrow-band detector.

In contrast to astrophysical electro-magnetic radiation, all 
GW signals, with the exception of stochastic 
backgrounds and signals of unknown shape, are tracked in phase and the 
SNR is built up by coherent superposition of many cycles emitted by
a source. In this case it is possible to deduce the detailed dynamics
of the bodies emitting radiation and a precise picture of the source 
can be reconstructed.
Since one tracks the phase of a signal, the SNR is
proportional to the amplitude and only falls off as $r^{-1}$ with distance.
Therefore, the number of sources accessible to a GW detector that is 
sensitive to sources with a limiting amplitude $h_{\rm min},$ increases as  
$h_{\rm min}^{-3},$ as $h_{\rm min}$ decreases.  Thus GW 
detectors can quickly improve their sensitivity and begin to see
sources at the edge of the Universe.
Astrophysical and cosmological GW have a frequency range of 
$10^{-4}$--$10^4$~Hz and the detectors are generally quite broad band.

\begin {figure}
\centering
{\epsfxsize 3.0in \epsfbox{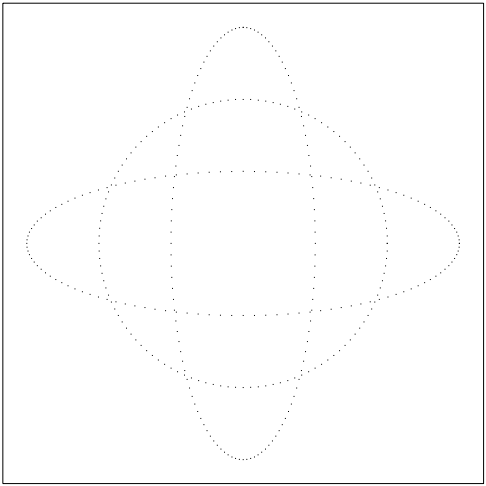} 
\epsfxsize 3.0in \epsfbox {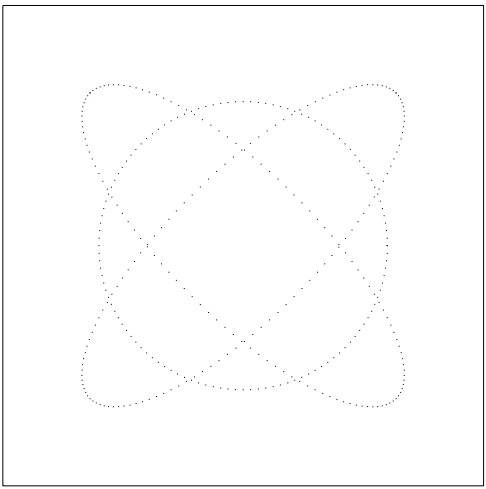}}
\caption {In Einstein's theory gravitational waves have two
independent polarisations. The effect on a circular ring of particles in the 
$(x,y)$ plane due to a plus-polarised wave traveling in the $z$-direction 
is shown in the left panel and due to a cross-polarised wave is shown in the
right panel. 
The ring continuously gets deformed into one of the ellipses and 
returns to the circular configuration during the
first half of a GW period and gets 
deformed into the other ellipse and back during the next half.}
\label{fig:polarisation}
\end {figure}

GR can be represented by a second rank, symmetric 
trace-free tensor.
In a general coordinate system, and in an arbitrary gauge, this tensor
has nine independent components. However, as in the electro-magnetic 
case, GR has only two independent states of 
polarisation in Einstein's theory: the plus-polarisation $h_+$ and the 
cross-polarisation $h_\times$ (the names being derived from the shape of the
equivalent force fields they produce). However, unlike electro-magnetic 
radiation the angle between the two polarisation states is
$\pi/4$ rather than $\pi/2.$ This is illustrated in
Fig.~\ref{fig:polarisation} where the response of a ring free particles
in the $(x,y)$ plane to plus-polarised and cross-polarised GW 
traveling in $z$-direction is shown. Measuring the polarisation states is 
of fundamental importance since there are theories of gravity, other than 
general relativity, in which the number of polarisation states is more 
than two; in some theories even dipolar and scalar waves exist \cite{will.81}. 
Polarisation measurement is a qualitative test: Even a single
measurement of a polarisation state that is not plus or cross,
can rule out general relativity. Such measurements would be 
possible with a network of detectors.  

Measuring the polarisation of the waves has astrophysical 
implications too. Astrophysical observations can only determine 
the mass of a binary to within a factor $\sin \theta$ where $\theta$
is the unknown inclination angle of the binary with the line-of-sight.
From GW polarisation measurements of astrophysical 
binaries we will be able to infer the inclination angle which will
further help us to resolve the mass-inclination angle degeneracy.

\begin {figure}
\begin{center}
\centering
\epsfxsize 3.0 in \epsfbox{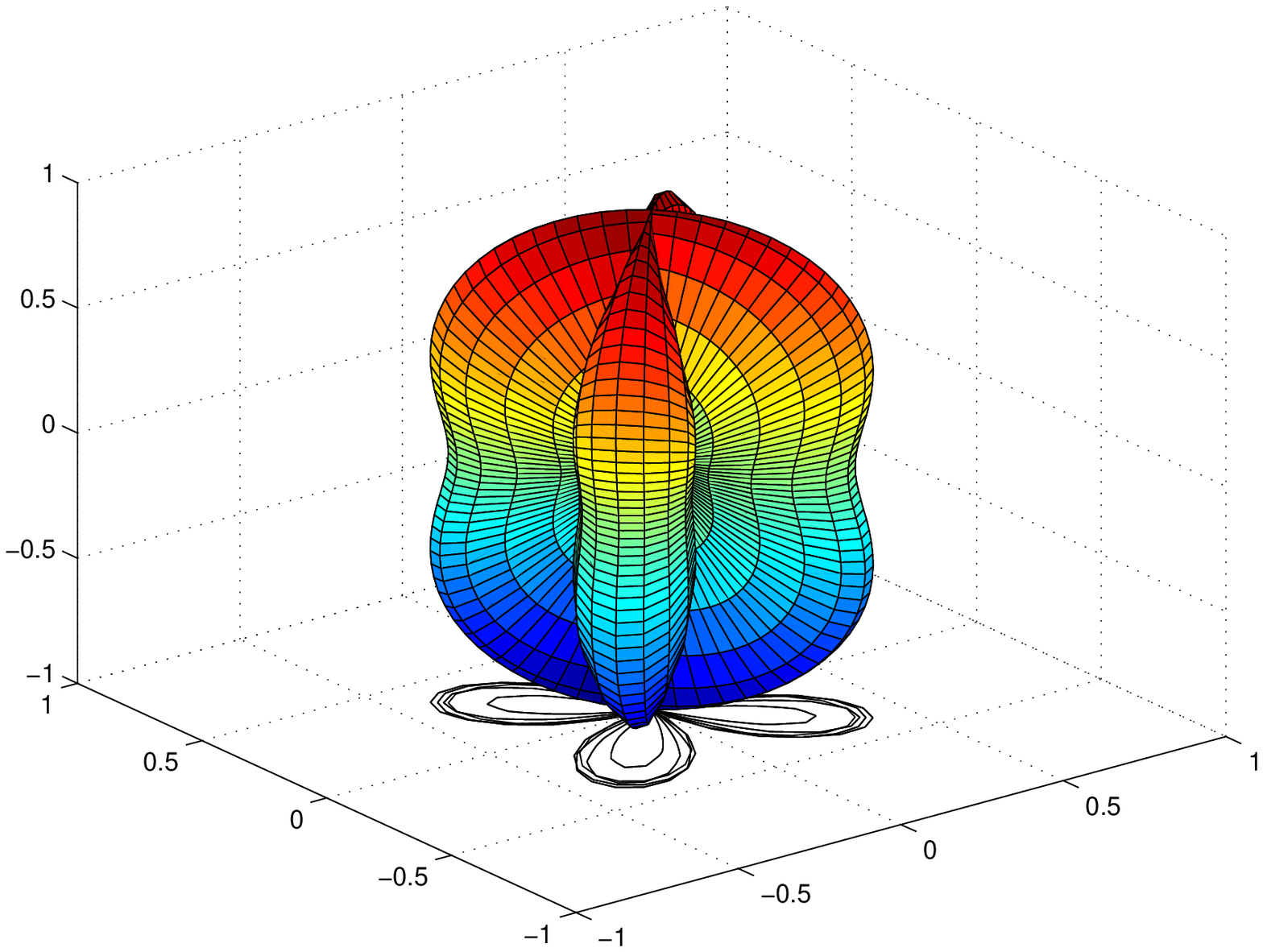}
\epsfxsize 3.0 in \epsfbox{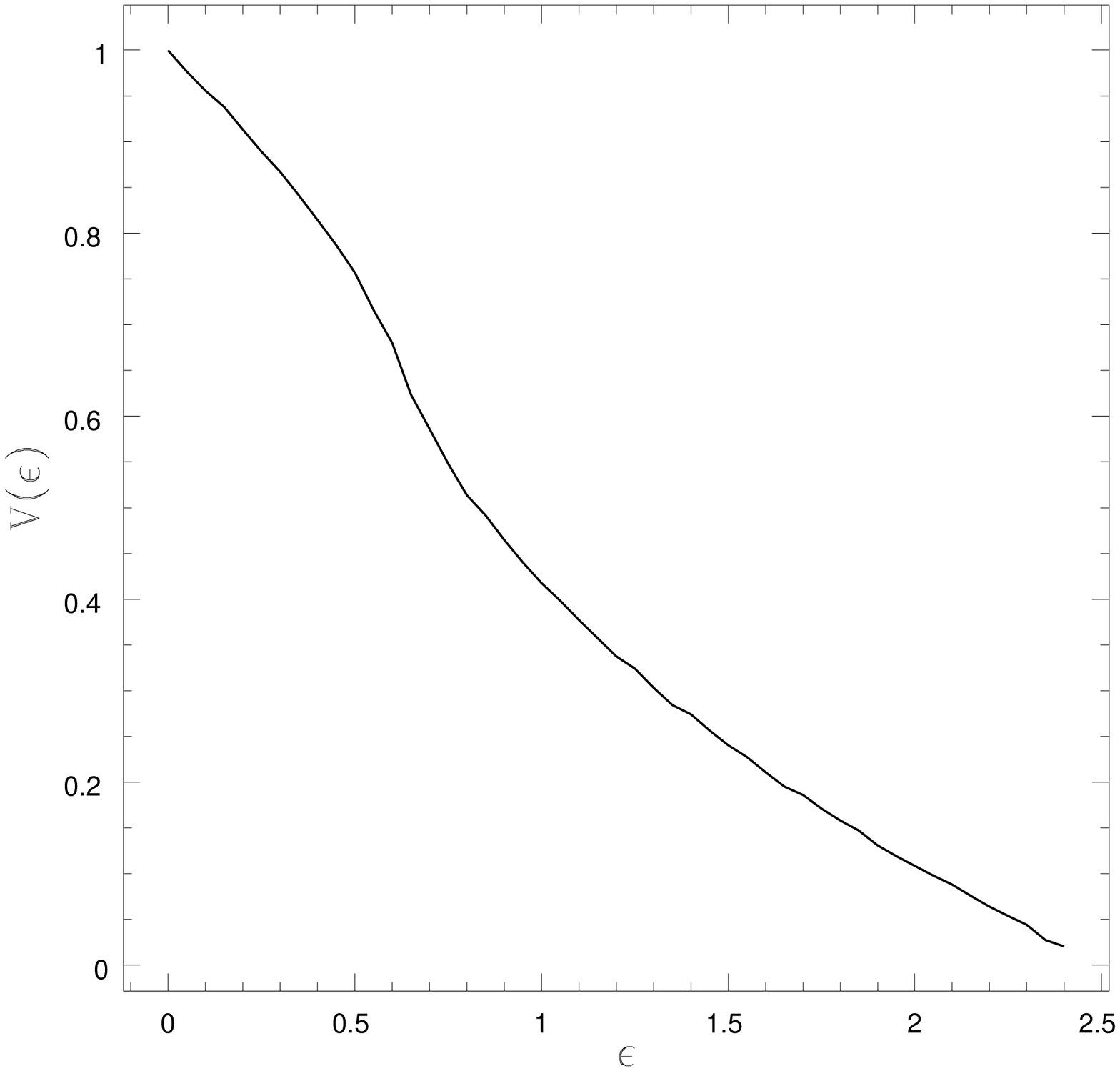}
\caption {The antenna pattern of an interferometric detector (left panel).
The pattern is shown for a circularly-polarised waves. The response for 
a wave traveling in a certain direction 
is proportional to the distance from the origin, at the centre of the
plot, to the point on the antenna pattern 
in that direction.  Also shown is the fractional area on the sky over which 
the response 
is a certain fraction $\epsilon$ of the average (right panel). The average
response has been arbitrarily set to unity.}
\end{center}
\label{fig:response}
\end {figure}

\section{Gravitational wave detectors and their sensitivity}
\label{sec:detectors}

Our detectors cannot directly measure the two independent polarisations 
of the waves but rather a certain linear combination of the two.
The complex response $R$ of a detector is 
\cite{schutz.tinto,dt88}:
\begin{equation}
R = h_+ F_+ + i h_\times F_\times,\ \ F_+ = \Re(F),\ \ F_\times=\Im(F),
\end{equation}
where $F_+$ and $F_\times$ are the real and imaginary parts of the complex
antenna pattern, which is a function of the direction 
$(\theta,\phi)$ to the source and the polarisation angle $\psi$
of the wave, in a coordinate system `attached' to the detector. Responses
of three widely separated detectors to an incident GW,
together with two independent differences in arrival times among them, 
are, in principle, sufficient to fully reconstruct the GW. For a resonant bar detector, with its longitudinal axis
aligned with the $z$-axis, the response is,
\begin{equation}
F=\sin^2\theta e^{2i\psi},
\end{equation}
and for an interferometer with its arms in the $(x,y)$ plane and
at right angles to each other and the $x$-axis bisecting the two arms,
\begin{equation}
F= \left [\frac{1}{2}\left (1+\cos^2\theta \right ) \sin 2\phi +
i \cos\theta \cos 2\phi  \right ]
e^{2i\psi}.
\end{equation}

In the following we shall assume that the two polarisation amplitudes
are equal: $h_+=h_\times \equiv h/\sqrt{2},$ that is circularly
polarised waves.  The response of an interferometric detector is plotted in 
Fig.~\ref{fig:response}.  The rms response of an interferometric detector,
averaged over the entire sky, is smaller 
by a factor of $\sqrt{5}$ than the maximum response.
Fig.~\ref{fig:response} also shows the percentage area of
the sky over which the response in an interferometric antenna
is larger than a certain fraction $\epsilon$ of the rms response.  
The response is better than 50\% of the rms
over 75\% of the sky, implying that interferometric detectors are
quite omni-directional. In contrast, the rms response of a resonant bar
is smaller by a factor of $1/\sqrt{15}$ than the optimum response and the
response is better than 50\% over 69\% of the sky. Thus, bars too are
almost as omni-directional as interferometers.

Currently, there are a number of  bar detectors in operation: some
of these operate at room-temperature and some others at cryogenic 
temperatures. Bar detectors are resonant, narrow-band detectors.
They can detect
signal amplitudes $h \sim 10^{-20}$ in a band 
width of 10--20~Hz around a central frequency that lies at about 1 kHz. 
Asymmetric supernovae in our Galaxy are the best candidate sources
for these detectors. They may also see continuous radiation emitted 
by a neutron star if the frequency happens to lie in their 
sensitivity band.

Interferometric detectors currently under construction will 
increase our ability to directly observe GW. 
The Japanese have already built a 300 m detector 
TAMA in Tokyo, Japan~\cite{tama}.  
Several other projects are now nearing completion:
The British-German collaboration is 
constructing a 600 m interferometer GEO in Hannover, Germany \cite{geo}, 
the French-Italian collaboration is building a 3 km detector 
VIRGO near Pisa, Italy \cite{virgo} and the Americans are
building two 4 km antennas LIGO, one in Livingston and the other 
in Hanford \cite{ligo} in the U.S.A. These detectors 
will start taking data between 2001 and 2003. The larger of these
detectors, LIGO and VIRGO, are likely to  be upgraded in sensitivity
by an order of magnitude with a better low-frequency performance
in 2006.  These ground based interferometers will eventually be
sensitive to sources in the frequency range from 1 Hz to several kHz.

One has to move to space to observe GW sources whose time periods are larger
than about 1 second. Currently the prospect for constructing the
Euro-American Laser Interferometer Space Antenna (LISA) looks very 
good\cite{lisa}.  LISA consists of three drag-free
satellites, forming an equilateral triangle of size 5 million km 
and in an heliocentric orbit, lagging behind Earth by 30 degrees.
LISA will be sensitive to waves in the low-frequency band of 
$10^{-4}$--$1$~Hz.

\begin{figure}
\centering
\epsfxsize 3.0 in \epsfbox{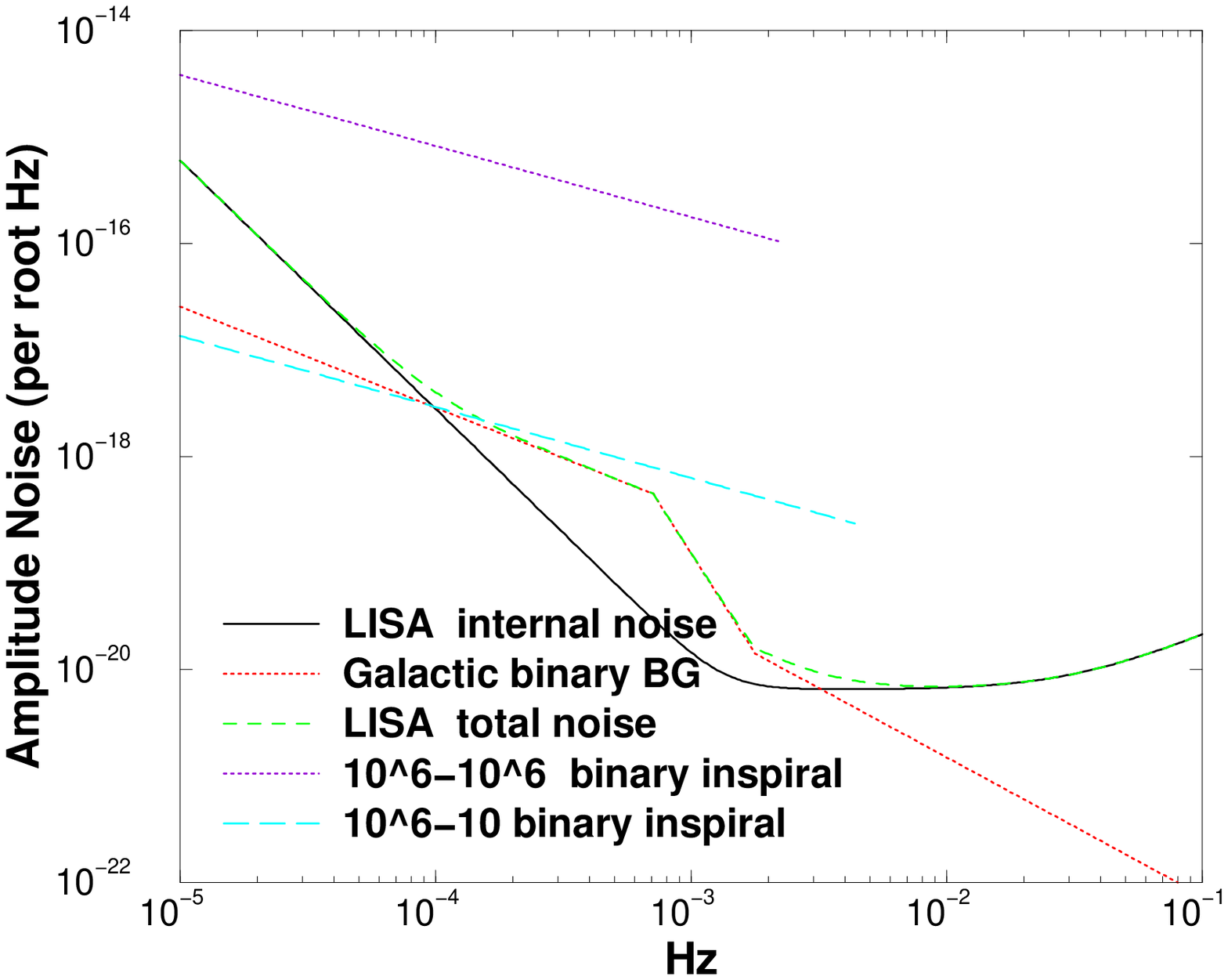}
\epsfxsize 3.0 in \epsfbox{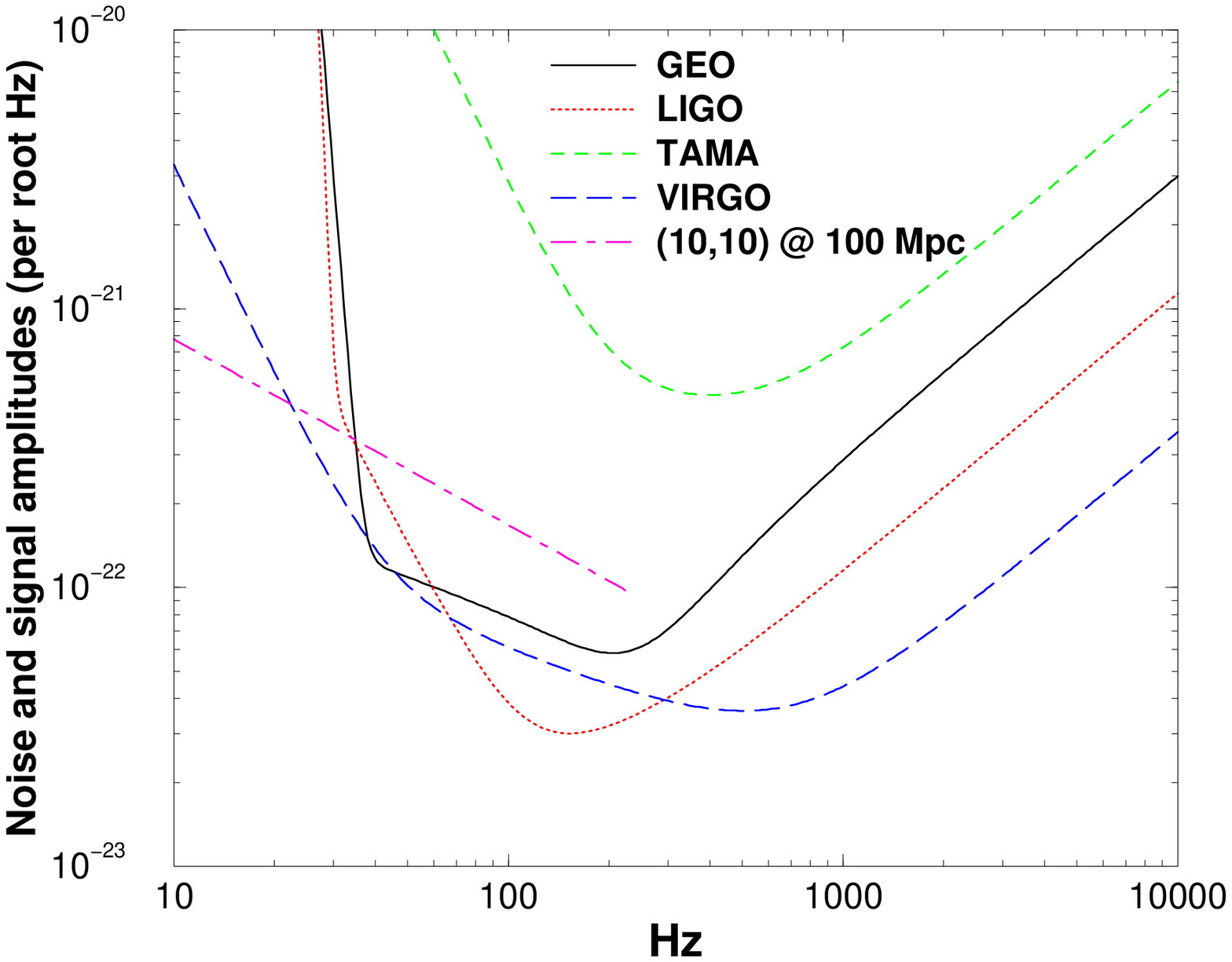}
\caption{The amplitude noise spectral density,
$\sqrt{S_h(f)},$ in ground- and space-based interferometers.
On the same graph we also plot the signal 
amplitude, $ \sqrt{f\,}|\tilde h(f)|$, of binary black holes --
inspiral occurring at a distance of 100 Mpc in the case of stellar mass 
black holes and at 3 Gpc in the case of super-massive black holes. 
LISA can super-massive black hole mergers with a good SNR 
practically anywhere in the Universe.}
\label{fig:noise curves}
\end{figure}

The performance of a GW detector is characterized by the 
{\it power spectral density} (PSD) $S_h(f)$ of its noise
background which has dimensions of Hz$^{-1}$ (see \cite{grishchuk et al}
for a discussion on the noise PSD of interferometers).  
One compares the GW amplitudes of astronomical 
sources with the instrumental sensitivity and assesses
what sort of sources will be observable in the following way:
The Fourier component $\tilde h(f)$
of a deterministic signal $h(t)$ has dimensions of Hz$^{-1}$ and 
the quantity $f |\tilde h(f)|,$ is dimensionless. This is 
compared with $h_n(f)\equiv fS_h(f),$ the dimensionless
noise PSD per logarithmic bin, to deduce the strength of a
source relative to the detector noise.  
The noise PSD of ground- and space-based interferometric GW
detectors is shown in Fig.~\ref{fig:noise curves}.  

It is quite common also to 
compare a signal's amplitude spectrum per logarithmic bin, 
$\sqrt {f} |\tilde h(f)|,$ with the amplitude spectrum of 
noise, $\sqrt {S_h(f)},$ both having dimensions of per 
root Hz.  Finally, for monochromatic sources, one compares the 
effective noise in a long integration period with the 
expected ``instantaneous'' amplitudes as follows:
A monotonic wave of frequency $f_0$ observed 
for a time $T$ is simply a narrow line in a frequency bin
of width $\Delta f\equiv 1/T$ around $f_0.$ The noise in this
bin is $S_h(f) \Delta f=S_h(f)/T.$ 
Thus the SNR $\rho$ after a period of observation $T$ is 
$\mbox{SNR} = h_0/\sqrt {S_h(f_0)/T}.$
One therefore computes this dimensionless noise spectrum for
a given duration of observation, $S_h(f)/T,$  to assess the 
detectability of  a continuous GW. 

\section {Sources of gravitational waves}
\label {sec:sources}
Inferences drawn from astronomical observations have led to the belief that 
luminous matter constitutes
a small fraction of the total matter content of the Universe. More
than 90\% of the mass in the Universe is electro-magnetically silent. The 
presence of dark matter is inferred from the gravitational influence it
causes on luminous matter. It is possible that some fraction
of this dark matter is a strong emitter of GR.
Today our firm sources are only those inferred via astronomical observations
but it is only when we learn how to make gravitational observations of these
sources can we hope to explore the unknown Universe in detector
data. The reason for this is partly because we do not know what sort 
of radiation to expect from these plausible sources 
and partly because GW data analysis is a complex process 
involving 10's of Terabytes of data each year, from hundreds
of different detector channels. After an initial steep learning curve
gravitational observations could prove to be a very rewarding exercise. 
In this Section we will take a look at some of the anticipated sources.

\subsection {Gravitational wave bursts}
GW carry energy and momentum from their sources. By 
demanding that the energy lost in the form of GR is 
precisely balanced by the decrease of the energy in the system, one can derive
a simple expression for the apparent luminosity of radiation ${\cal F},$
at great distances from the source, in terms of the 
GW amplitude \cite {schutz.1985}:
\begin {equation}
{\cal F} = \frac {| \dot h |^2}{16 \pi},
\end {equation}
where $\dot h$ denotes the time-derivative of the GW amplitude $h$.
This equation can be used to make order-of-magnitude estimates of 
GW amplitudes. 
If a source at a distance $r$ radiates energy $E,$ at a frequency $f,$
in a time $T,$ then writing $\dot h= 2\pi f h$ and ${\cal F}=E/(4\pi r^2 T),$
the amplitude is 
\begin {equation} 
h \sim \sqrt{ \frac{E}{T} } \frac{1}{\pi rf }.
\label {eq:amplitudeA}
\end {equation}
The amplitude of a fiducial stellar mass source that deposits a small 
fraction of its total mass in the form of GW over
a time scale of a few milliseconds would be 
\begin {equation}
h \sim 4 \times 10^{-21}
\left ( \frac{E} {10^{-7} M_\odot }  \right )^{1/2}
\left ( \frac{5~\rm ms}  {T}\right )^{1/2}
\left ( \frac{\rm 200~Hz}{f}  \right )
\left ( \frac{40~{\rm kpc}}{r}  \right ).
\label {eq:amplitudeB}
\end {equation}
Two masses separated by a distance of $\ell =1$~km will be tidally distorted
by no more than $4 \times 10^{-18}$~m by such a wave. This is a change in
length that is smaller than the size of a neutron and a phase change
of order $5 \times 10^{-10}$ radians in an interferometer that uses
0.1 micron laser and whose arms are 4 km long.

Long baseline interferometers, such as LIGO and VIRGO, aim towards
achieving a dimensionless noise amplitude $\sqrt{f S_h(f)}$ 
of $4 \times 10^{-22}$  
at about 200 Hz during their initial operation. 
From Eq.(\ref {eq:amplitudeB}) we see that an event at 
40 kpc, a distance covering our own Galaxy and some of its satellites,
depositing energy equivalent to $10^{-7} M_{\odot}$ at 200 Hz 
in 5 ms will appear in these detectors with a maximum SNR
of 10 or a sky and polarisation averaged SNR (which is $\sqrt{5}$ smaller
than the maximum) of about 4.5.  This is quite a good SNR for a network
containing 3 or more interferometers.

The numbers we have used for energy, duration and frequency are 
typical of what is expected for a strong type II supernova.
It is expected that an optically bright supernova event may occur as often as 
once every 30 to 100 years per galaxy. If for some reason gravitational collapse
and the ensuing supernova, in a good fraction of these events, are not optically
bright then the number of gravitationally observable events might 
be somewhat higher and within the region which the initial interferometers 
will be able to survey there may be several events per year. 
In any case second generation antennas should indeed observe quite a large
number of these events shedding some light on the nature of the gravitational
collapse. Future detectors may give us clues on the equation
of state of dense nuclear matter of gravitationally collapsed stars by
observing the radiation emitted as the collapsed star settles
downs to a quiescent state.

\subsection {Matched filtering and effective amplitude}

Theoretical calculations sometime allow us to accurately predict the
evolution of a system and the phasing of the GW that it emits. 
When the time development of a signal is known, one can cross-correlate 
the detector output with a copy of the expected signal. The process 
of cross-correlation leads to an enhancement in the SNR 
in proportion to the square-root 
of the number of cycles the signal spends in the detector band.  Note,
however, that one will have to perform the cross-correlation of the
detector output with a number of different copies of the signal, each
corresponding to a different set of values of the signal parameters; this
process is called {\it matched filtering.} In spirit, matched filtering 
of arbitrary signals is the same as a simple Fourier transform of
mono-chromatic signals.  For illustrative purposes, therefore, we shall
assume that the signal is mono-chromatic. A signal of frequency
$f$ lasting for a time $T$ would produce $n=fT$ cycles. Using this
in Eq.~(\ref{eq:amplitudeB}) we can define
an effective amplitude $h_{\rm eff}\equiv \sqrt {n}\times h:$ 
\begin {equation}
h_{\rm eff} = 4 \times 10^{-21}
\left ( \frac{E} {0.05 M_\odot }  \right )^{1/2}
\left ( \frac{\rm 200~Hz}{f}  \right )^{1/2}
\left ( \frac{30~{\rm Mpc}} {r} \right ),
\label {eq:eff.amplitude}
\end {equation}
where we have used the energy and frequency appropriate for a stellar
mass black hole binary close to its coalescence. From the above equation
we conclude that matched filtering helps in detecting signals as long as the 
effective amplitude $h_{\rm eff}$ is larger, even though the 
instantaneous amplitude $h$ could be much smaller, than the 
noise background. Indeed, the instantaneous amplitude can be smaller 
than the minimal detectable amplitude of a detector by a factor of $\sqrt{n}.$
The number of cycles for supernovae, black hole quasi-normal modes 
and other impulsive events is of order 10, for binary 
inspiral coalescences it can vary between $10$ to $10^{4}$
and for spinning neutron stars in a year's observation it 
can be as large as $10^{10}.$
Matched filtering can, therefore, greatly improve the visibility of
signals of known shape. This is the reason why there is a lot of emphasis
on studying detailed dynamics of GW sources. Of all sources, black hole
binaries and normal modes of neutron stars have been studied in 
greatest detail. 

\subsection {Binary black holes and neutron stars}

Binaries consisting of double neutron stars (NS), double black holes (BH) 
or a NS and a BH, are called compact binaries. Compact binaries are among the
best studied and most promising sources for the upcoming interferometric
GW antennas. In this Section we will discuss the prospect for detecting these
systems.
\subsubsection {The evolution of a compact under radiation reaction}
Gravitational waves carry energy and angular momentum from a 
compact binary causing the system
to eventually merge into a single black hole, emitting a burst of radiation
in the process. The amplitude and frequency of the waves
increase as the binary inspirals resulting in a characteristic chirping
signal. Post-Newtonian (PN) theory has been used to 
compute the evolution of such a system
but as the two stars approach each other the PN theory becomes less and 
less accurate, until finally, when the two stars are at a distance 
$r\simeq 6M,$ the PN approximation completely breaks down. 

Recent advances in radiation reaction calculations 
\cite {binary radiation reaction} and PN theory \cite{binary inspiral PN}
and their refinements \cite{DIS1}, have shed some light on 
the merger phase \cite{BD1 and 2}.  While these developments will help to some 
extent in tracking the inspiral and coalescence, a more 
careful treatment of the problem using the full Einstein equations
will be needed to follow the non-linear evolution that takes place during
the merger phase. The radiation emitted at very late times --
called quasi-normal modes -- is again well understood using black hole
perturbation theory. The missing gap is the merger phase which could
last several 10's of milli-seconds in the case of stellar mass binaries. The
problem in the case of a stellar mass black hole falling into a super-massive
black hole is possibly extremely complicated owing to the presence of
spinning bodies and eccentric orbits, and we may have to wait 
several years before any useful solutions are found.

\subsubsection {White-dwarf binary inspiral}
\label{sec:whit-dwarf inspiral}

Ground based detectors are not sensitive to non-compact binaries, such as a 
white dwarf-white dwarf pair, which do emit GW, but at a
much lower frequency than where these detectors are sensitive.
GW amplitude would be the greatest when
the two component stars are closest. For a pair of 
stars each of radius $R,$ the distance of 
closest approach is $r=2R.$ The
Keplerian frequency at this distance is
\begin{equation}
f_{\rm orb} = \frac {1}{2\pi}\sqrt{\frac{M}{r^3}} = 0.1 
\left ( \frac{M}{M_\odot} \right )^{1/2}
\left ( \frac{r}{6000~\mbox{km}} \right )^{3/2} \mbox{Hz},
\end{equation}
where $M$ is the total mass of the system and the radius is taken
to be that of a typical white dwarf, $R=3000$~km.  We see that 
the largest GW frequency, which is equal to twice the orbital frequency, 
is outside the sensitivity band of
ground-based detectors. Indeed, much before the white-dwarfs begin to touch
each other the tidal interaction becomes important so that only frequencies
ten to hundred times smaller than the above limit can be observed.
Such frequencies are within the sensitivity band of LISA. In fact,
the Galactic and extra-galactic populations of white-dwarf binaries create a
stochastic GW background as shown in Fig.~\ref{fig:noise curves} 
and only the nearest sources will stand above the background noise.

\subsubsection {Compact binary inspiral}
\label{sec:compact inspiral}

For compact binaries the adiabatic inspiral phase, when the time-scale
for frequency evolution is significantly larger than the orbital time-scale,
ends well before the two stars begin to touch each other. This is because the
non-linear general relativistic effects will alter the dynamics of the two
bodies making the adiabatic approximation and the PN theory invalid. 
Apart from the break down of the physical assumptions, it is, indeed, 
even technically impossible to continue the orbital evolution beyond 
the last stable orbit under the adiabatic approximation \cite{DIS3}. 
For a test particle in Schwarzschild geometry the last stable circular 
orbit (LSO) occurs at a distance $R_{\rm lso}
= 6M,$ corresponding to a LSO GW frequency $f_{\rm lso}$ given by
\begin{equation}
f_{\rm lso} = \frac {1}{6^{3/2}\pi M} = 
1.5 \left ( \frac {M}{2.8~M_\odot} \right)^{-1}~{\rm kHz}r.
\label{eq:lso}
\end{equation}
Binaries more massive than a typical NS binary of
$M=2.8 M_\odot$ end their inspiral phase at lower frequencies. 
The mass of a binary whose LSO frequency is the same as the
frequency $f_{\rm peak}$ at which a detector has the 
best effective sensitivity is given by 
\begin{equation}
M_{\rm optimal} 
= 15 \left (\frac {f_{\rm peak}}{150~{\rm Hz}}\right )^{-1} M_\odot.
\end{equation}
Therefore, stellar mass binaries are best suited for ground-based detectors.
Astronomical observations have evidence for the existence of stellar mass
single black hole systems with their mass in the range 
5-15 $M_\odot.$ However, there has been no observational evidence for 
a double black hole system. There is a real
difficulty in observing such systems astronomically as they cannot be expected
to have an accretion disc, which is the source of electro-magnetic radiation
in single black hole systems. Ground-based GW 
detectors should eventually shed light on the occurrence of black hole
binaries in nature.

LISA has its peak sensitivity $f_{\rm peak} \simeq 1.5 \times 10^{-3}$~Hz
implying that binaries with a total mass $M=1.5 \times 10^{6}M_\odot,$ 
are the best candidate sources. There is mounting astronomical evidence 
that the centre of every galaxy hosts a massive black hole of mass in the range
$10^5$-$10^9M_\odot$ depending on the type of galaxy at hand \cite{bhevidence}.
Moreover, images of the high red-shift Universe taken by the Hubble space
telescope shows that galaxies were strongly interacting in the past
and that most galaxies must have undergone merger at least once
in their life-time. This merger would have resulted in the coalescence
of the black holes at their centres. Moreover, the dense environs of galactic
centres also host many stellar mass compact objects that are captured
by the central only once in a while leading to the growth of the hole's mass.
LISA will be able to
detect GWs emitted in the process of massive BH inspiral and merger
wherever they occur in the Universe as well as stellar mass   
compact objects sinking into, and coalescing with, massive black holes.

The duration of an inspiral signal observed by an antenna depends on the
masses of the component stars and the lower frequency cutoff of the detector
sensitivity window. An antenna with a lower cutoff frequency $f_{\rm L}$ and
upper frequency cutoff larger than the LSO frequency will
observe the signal for a time $\tau$ given by 

\begin {equation}
\tau = \frac {5}{256 \eta M^{5/3} (\pi f_{\rm L})^{8/3}} 
\left [ 1 - \left (\frac{f_{\rm L}}{f_{\rm lso}} \right )^{8/3}\right ], \ \ \
f_{\rm L} < f_{\rm lso}.
\end {equation}
If $f_{\rm L} \ll f_{\rm lso}$ then
\begin {eqnarray}
\tau & \simeq & 24.2 \left ( \frac {0.25}{\eta} \right )
\left ( \frac {2.8 M_\odot}{M} \right )^{5/3}
\left ( \frac {40~{\rm Hz}}{f_{\rm L}} \right )^{8/3}~{\rm s} \nonumber \\
& \simeq & 1 \left ( \frac {0.25}{\eta} \right )
\left ( \frac {2 \times 10^6 M_\odot}{M} \right )^{5/3}
\left ( \frac {45.4 ~{\rm \mu Hz}}{f_{\rm L}} \right )^{8/3}~{\rm yr},
\label{eq:chirptime}
\end{eqnarray}
where $M=m_1+m_2$ is the total mass and $\eta=m_1m_2/M^2$ is the symmetric
mass ratio.  For most binaries the signal will last
long enough to observe only once. In other words, all inspiral signals 
in ground-based detectors, and most in LISA, are transient events and the
question arises how many such events are we likely to observe in a certain
period of time.

Estimates of rates of binary coalescences are based either on the few 
observed NS-NS binaries in our Galaxy or theoretical studies of compact
binary formation and evolution on  a computer -- the so-called {\it population
synthesis.} Both these estimates are highly uncertain. The estimates
give a rate between $10^{-6}$--$10^{-5}$ NS-NS coalescences per year in 
each galaxy of size the same as our own Milky-way. This gives several
NS-NS coalescences per year in a volume of (300 Mpc)$^3.$ Black hole
binaries may have similar coalescence rate densities. However, being more 
massive than NS-NS binaries, BH-BH binaries can be observed from greater
distances and hence the number of expected BH-BH coalescences in initial
interferometers is greater.

Coalescence rates involving super-massive black holes is also quite uncertain.
But it is expected that each year LISA will observe 10-100 stellar mass
black holes falling into super-massive black holes and several
galaxy mergers. Formation of massive black holes at galactic centres has
not been well understood. Observations cannot completely rule out or favour
any one formation scenario. In most scenarios it should be possible to
observe the GW radiation emitted in the process of formation of the hole.
The newly formed black hole would emit radiation at a frequency characteristic
of its mass and spin.

During the inspiral regime the signal will sweep as many as 
\begin {equation}
N_{\rm cyc} = \frac {1}{32\pi \eta\left (\pi M f_{\rm L} \right )^{5/3}}
\simeq 1600 \left ( \frac {2.8 M_\odot}{M} \right )^{5/3}
\left ( \frac {40~{\rm Hz}}{f_{\rm L}} \right )^{5/3}
\end {equation}
cycles.  The energy emitted in the inspiral phase is equal to the binding
energy of the system at a separation $R=6M:$ 
\begin {equation}
E=-\frac {m_1 m_2}{2R} = -\frac {\eta M^2}{2R} = -\frac {\eta M}{12}.
\label {eq:binary.energy}
\end {equation}
Thus, in the case of equal mass binaries (i.e., $\eta=1/4$) the maximum 
energy emitted during the inspiral phase
is about $1/50$ of its total mass or roughly $0.05 M_\odot$ for a NS-NS binary,
which is what was used in equation (\ref{eq:eff.amplitude}). 
The effective amplitude at a given frequency $f$ can be obtained by noting
that the distance between the two stars is related to the GW
frequency via $R=[M/(\pi^2 f^2)]^{1/3},$ and using $|E|=\eta M^2/(2R) =
\eta M^{5/3}(\pi f)^{2/3}/2,$ in the expression for the effective amplitude,
$h_{\rm eff}=\sqrt{E/f}/(\pi r):$
\begin {equation}
h_{\rm eff} = \frac {\eta^{1/2} M^{5/6} f^{-1/6}}{\pi^{2/3} r}.
\end {equation}
The above equation gives,
\begin {eqnarray}
h_{\rm eff} & = & 3 \times 10^{-21} 
\left ( \frac{30~{\rm Mpc}}{r} \right )
\left ( \frac{\eta}{0.25} \right )^{1/2}
\left ( \frac {M}{2.8 M_\odot} \right )^{5/6}
\left ( \frac {1~{\rm kHz}}{f} \right )^{1/6}\nonumber \\
& = & 4 \times 10^{-17} 
\left ( \frac{1~{\rm Gpc}}{r} \right )
\left ( \frac{\eta}{0.25} \right )^{1/2}
\left ( \frac {M}{10^6 M_\odot} \right )^{5/6}
\left ( \frac {1~{\rm mHz}} {f}\right )^{1/6}.
\label{eq:eff.amplitude.binary}
\end {eqnarray}
The SNR for inspiral events in various
ground- and space-based interferometers are plotted
for stellar mass and super-massive sources
located at a fixed distance of 100 Mpc and 3 Gpc from Earth, respectively.
The initial network of ground-based antennas have a fair chance of making
the first observations of GW by tracking the inspiral
signals from stellar mass black holes, while LISA can observe neutron stars and
small black holes falling into super-massive black holes from up to a
red-shift of 1. Galaxy mergers and the associated
super-massive black hole coalescences will appear in LISA
as very luminous events wherever they occur in the Universe.
\begin{figure}
\centering
\epsfxsize 3.0 in \epsfbox{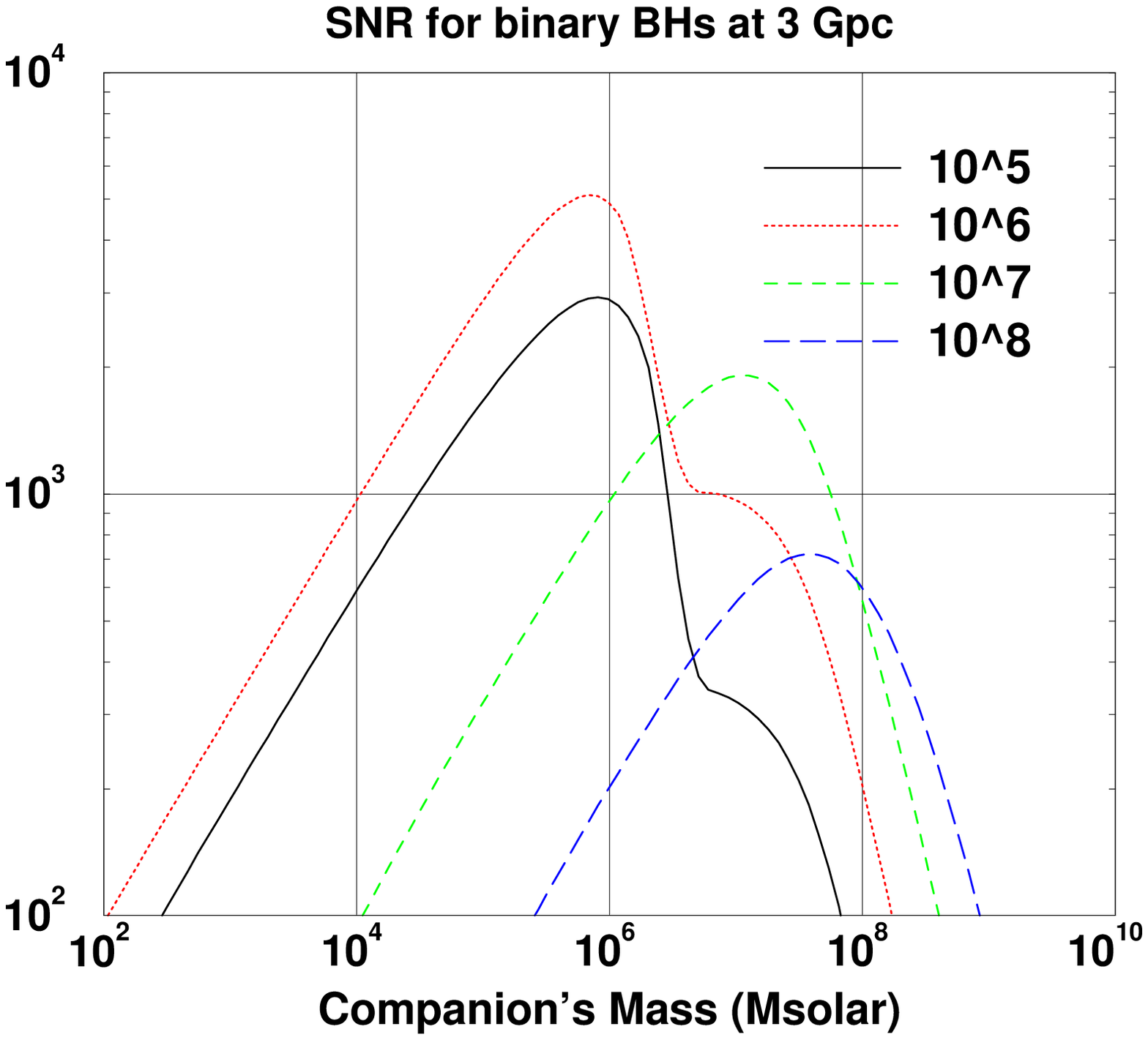}
\epsfxsize 3.0 in \epsfbox{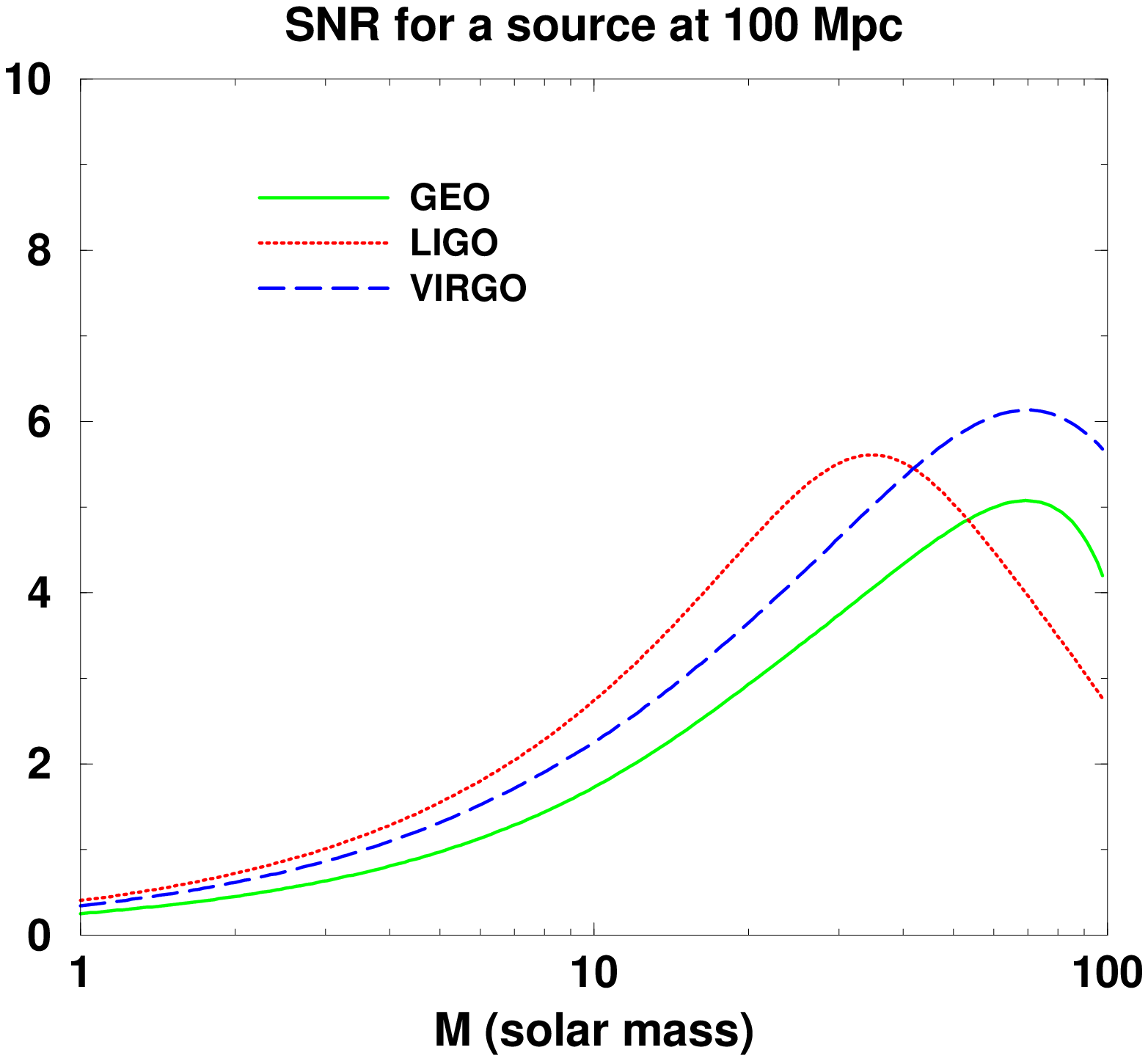}
\caption{The expected SNR from an inspiral event is plotted for ground-
and space-based interferometers. In the left hand panel we show the SNRs
as a function of the companion's for different masses of central black 
hole. For ground-based interferometers we show the SNR as a function
of the total mass of equal mass binaries but for different detectors.}
\label{fig:snrs}
\end{figure}

\subsubsection{Binary black holes as a probe of physics, astrophysics and cosmology}

Inspiral events are wonderful test beds for
general relativity, for both its fundamental predictions as well as
for testing the theory in the strongly nonlinear regime.  
The post-Newtonian expansion of the inspiral signal
has unfolded a rich cache of non-linear physics that governs the
dynamics of a binary. There are many interesting effects such as 
GW tails, spin-spin interaction, Lens-Thirring effect,
etc., which can all be deciphered 
from a high SNR event. For instance, the effect of 
tails on the phasing of GW can be detected in 
the inspiral waves emitted by BH-BH binaries \cite{lbbs95}.

Following an inspiral event with electro-magnetic
observation, such as gamma- or X-ray observation, 
would help to deduce the speed of GW to a phenomenal accuracy.
(Even a day's delay in the arrival times of gravitational and 
electro-magnetic radiation from a source at a distance of million light years
would determine the relative speeds to better than one part in
$10^{8}$).  This will require a good timing accuracy to determine 
the direction to the source so that astronomical observations of 
the same source can be made.  

One can tighten the limit on the mass of the graviton by observing the 
inspiral waves.  Massive gravitons would alter the phasing of the 
signal. Matched filtering facilitates an accurate measurement of the phasing 
which in turn allows to place a bound on graviton's
mass. Will estimates that one can bound the mass of the graviton to
$2.5 \times 10^{-22} $ eV using ground-based detectors and
$2.5 \times 10^{-26}$ eV using space-based detectors \cite {will.98}.

Inspiral waves are standard candles in the sense that by 
measuring their luminosity we can infer the distance to source
Schutz \cite{schutz.nature}.  In the quadrupole approximation, the
wave's amplitude (as well as its phase) depends only on a certain combination 
of the masses called the {\em chirp mass,} ${\cal M}=\eta^{3/5} M,$ and
the distance $r$ to the source. If we can measure the amplitude and the chirp 
mass independently, then we can infer the distance to the source directly.
Inspiral waves are detected by using matched filtering technique 
which enables an accurate estimation of the masses of the component stars.
In addition, the SNR enables us to measure the 
signal's amplitude as well. Thus, inspiral waves serve as standard candles.
This will allow us to determine the Hubble constant very accurately
by identifying the host galaxy and measuring its red-shift. It should
be noted that the inspiral wave's phase is quite sensitive to
post-Newtonian corrections and when they are included the mass-degeneracy
is lifted. It is, therefore, possible to determine both the masses of a binary
by using post-Newtonian search templates. 

Observations of massive black hole coalescences at cosmological distances 
$z\sim 1$ by space-based  detectors can facilitate an accurate determination 
of the distance to the source. Space-based detectors observe a massive binary
inspiral for a whole year and they have, thus, the baseline of the Earth's 
orbit around the Sun to triangulate the source on the sky. They can do this
to an accuracy of about a square degree. At a distance of 3 Gpc one square
degree is roughly 100 Mpc$^2$, a scale over which no more than one virialised
galaxy cluster can be found. Thus, an optical identification of the
host galaxy cluster and its red-shift should be possible, which would 
enable the measurement of the deceleration parameter $q_0,$ and 
hence the density parameter $\Omega,$ of the Universe. A single 
source at $z\gsim 1$ is enough to measure both $H_0$ and $\Omega$ 
to an accuracy of better than 1\%. 

Galaxy interactions and mergers must have been quite a common event in
the early history of galaxy formation. We know very little about the
history of galaxy formation and have no clue how super-massive
black holes may have formed. LISA will be able to detect galaxy
mergers practically anywhere in the Universe thereby helping in resolving
many puzzling questions about super-massive black holes.

By measuring GWs emitted during the inspiral of a stellar
mass compact object into a super-massive black hole we will be able to map the
spacetime geometry around massive, rapidly spinning holes and test 
the uniqueness of the Kerr solution, observe precessions of the 
periastron and of the orbital plane and associated non-linear effects. 
Such observations will undoubtedly help us understand the 
complicated dynamics and possibly even chaotic behaviour, expected 
in the evolution of such systems.
Thus, space-based detectors can potentially contribute quite a lot to 
further our understanding of fundamental science and cosmology.

\subsection {Black hole quasi-normal modes}
\label{sec:ring down}

A compact binary coalescing as a result of gravitational radiation reaction 
would most likely result in the formation of a single black hole.
The newly formed hole will initially be somewhat non-spherical, and 
this dynamical non-sphericity will be radiated away in the form of 
GW.  The late time behaviour of this radiation is well studied in the
black hole literature and there are detailed calculations of the
(quasi) normal modes for both static, i.e.
Schwarzschild, and stationary, i.e. Kerr, black holes. In all cases
the time-evolution of the emitted radiation is well-modelled
by a quasi-periodic signal of the form
\begin{equation}
h(t; \tau, \omega) = A e^{-t/\tau} \cos(\omega t)
\end{equation}
where $\tau$ is the decay time-scale of the mode in question and
$\omega$ is the angular frequency of the mode, both of which depend
on the black hole's mass and spin. In all but the
extreme Kerr black holes (extreme Kerr black holes are those
that are spinning at the maximum possible rate) the only dominant
mode, i.e. the mode for which the decay time is the longest and
the amplitude is the highest, is the fundamental mode  whose
frequency is related to the mass $M$ and spin $a$ of the black hole
via $\omega=[1-0.63 (1-a)^{0.3}]/M$ where
$M$ is the mass of the black hole in units $G=c=1$ and $a=J/M$ is the
spin angular momentum of the hole in units of black hole mass. The decay
time $\tau$ is given by $\tau = 4/[\omega (1-a)^{0.45}].$ 
(See Ref.~\cite{Flanagan&Hughes98} and references therein for details.)

It is estimated  \cite{Flanagan&Hughes98} that 
during the quasi-normal mode ringing
of a black hole the energy emitted might be as large as 3\% of the 
system's total mass.  By matched filtering 
it should be possible to
detect quasi-normal modes, in initial interferometers, from black holes
of mass in the range 60--$10^3$M$_\odot$ and at a distance of 
200~Mpc.  Binary black hole mergers should result in the emission of
such a ring down signal during the late stages. Thus, inspiral signals
emitted before the merger might aid in identifying the quasi-normal modes.

\subsection {Continuous waves} 
\label {sec:periodic.sources}

Our Galaxy is expected to have at least $10^{8}$ spinning neutron
stars that form roughly at a rate of one every 30 years. Some 
population of neutron stars is in binaries. There are a number of 
ways in which a single spinning neutron star could radiate away 
GW (if the neutron star is axisymmetric, of course, 
then there will be no GW emission): (1) Neutron stars 
normally spin at high rates (several to 500 Hz) and this must induce
some equatorial bulge and flattening of the poles. The
presence of a  magnetic field may cause the star to spin
about an axis that is different from the symmetry axis
leading to a time-varying quadrupole moment. 
(2) The star may have some density inhomogeneities in the
core/crust set up during its formation and/or subsequent convectively 
unstable motions of the core. (3) The presence of an accretion disc, 
with its angular momentum not necessarily aligned with that of the 
neutron star, can potentially alter axisymmetry. That and electromagnetic 
radiation reaction torques can induce and sustain wobble. 
(4) The normal modes of the neutron star fluid  (radial and other
oscillations) can extract rotational energy and re-emit in the form
of GW. (5) There are certain classical and relativistic
instabilities in the neutron star fluid which may cause the star
to radiate away energy in the form of GR. In what
follows we will only discuss a sample of recent work on the radiation
from spinning neutron stars.

\subsubsection{GW amplitude from spinning asymmetric neutron stars}

If $I_{zz}$ is the moment of inertia about the spin axis of a neutron
star emitting GW at a frequency $f$ then the 
gravitational amplitude at a distance $r$ is:
\begin {equation}
h = 3 \times 10^{-27} \left ( \frac {10~{\rm kpc}}{r} \right )
\left ( \frac {I_{zz}} {10^{45}~g~cm^2} \right )
\left ( \frac {f}{200~{\rm Hz}} \right )^2
\left ( \frac {\epsilon}{10^{-6}} \right ),
\label{eq:amplitudeC}
\end {equation}
where $\epsilon$ is the ellipticity of the star. 
In a simple model of an equatorial plane of elliptical cross section with
semi-major axis $a_1$ and semi-minor axis $a_2,$ the ellipticity
is $\epsilon \equiv 1-a_2/a_1$. The ellipticity is an unknown but
one can obtain an upper limit on it by attributing the observed 
spin-down of pulsars $\dot P$ to gravitational radiation back reaction, 
namely that the change in the rotational energy $E=I\Omega^2/2$ 
is equal to GW luminosity. Then, the ellipticity  
is related to the spin-down rate of a pulsar via
\begin {equation}
\epsilon = 5.7 \times 10^{-6} 
\left ( \frac {P}{10^{-2}~{\rm s}} \right )^{3/2}
\left ( \frac {\dot P}{10^{-15}} \right )^{1/2}.  
\end {equation}
Since one knows the observed values of $P$ and $\dot P$ one can obtain
an upper limit on $\epsilon$ using the above equation. Following this
method one finds that for the Crab pulsar 
$\epsilon \le 7 \times 10^{-4}.$ The gravitational wave amplitude corresponding
to this ellipticity is $h \le 10^{-24}.$ Noting that Crab has a spin frequency
of 25 Hz (GW frequency of 50 Hz) equation 
(\ref{eq:eff.amplitude.binary}) implies an effective amplitude $h_{\rm eff}
= 4 \times 10^{-20}$ after a year's integration and hence 
observable in initial interferometers. \cite{Schutz 99}. 
It is unlikely that the ellipticity is so large and hence the effective
GW amplitude is probably much less.  Yet, the prospect
of seeing Crab at a tenth, or even a hundredth, of this ellipticity is
quite good with first/second generation interferometers.

\subsubsection{Relativistic instabilities in young neutrons stars}

It has long been known that the fundamental
mode of a neutron star fluid, undergoes unstable,
as a result of GW emission \cite{cfs}, when the spin frequency of the
star is larger than a certain critical spin frequency.
The physics behind this instability can be understood in the following 
manner: 
Imagine exciting a mass-quadrupole mode -- that is a non-uniform distribution
of mass -- in a non-spinning star. The mass inhomogeneity will travel on the
surface of the star at a fixed speed relative to the star and this 
dynamical asymmetry will cause the star to 
radiate GW. Gravitational waves drain the angular momentum 
from the modes as a result of which the modes will decay.  
Now consider a spinning neutron star in which a co-rotating and a 
counter-rotating modes are excited.  For spins smaller than the pattern 
speeds both these modes will decay in course
of time by losing angular momentum: co-rotating modes lose positive angular
momentum and counter-rotating modes lose negative angular momentum. 
But when the neutron star spins at a rate greater than a certain 
critical rate, to an external inertial observer 
both modes will appear to be co-rotating.
Therefore, the mode counter-rotating relative to the star will
also emit positive angular momentum, causing the angular momentum associated
with the mode to enhance, or for the amplitude of the mode to increase.
In other words, a mode counter-rotating relative to the star, but seen
co-rotating relative to the inertial observer, can only emit positive
angular momentum which causes its own angular momentum to increase.
The energy for this increase is supplied by the spin angular momentum
of the neutron star. Thus, while the mode co-rotating 
with the star's spin will decay, the mode 
counter-rotating with the spin will grow in amplitude and emit
more and more radiation. This will go on until the mode has sucked
out enough spin angular momentum of the star to make the counter-rotating
mode appear to be counter-rotating with respect to an inertial 
observer too. 

It is suspected that the CFS instability will not work in the presence of
viscosity and hence it may be unimportant in old neutron stars. However,
newly born neutron stars will be very hot and viscous forces may be 
insignificant in them.  Recently, another class of modes called 
$r$-modes have been discovered, \cite{andersson,lindblom et al,ak review} 
which -- unlike the CFS modes that are
mass-quadrupole moments -- are current-quadrupole moments that are unstable
at all spin frequencies of the star.
The physics of r-modes in neutron stars is rather
too complicated and there is as yet no definitive statement about the
role of these modes in GW emission in young neutron stars.

\subsection {Stochastic backgrounds}

Catastrophic processes in the early history of the Universe, as well as
the myriad astrophysical sources distributed all over the cosmos, generate
stochastic background of GW. Astrophysical backgrounds
carry the signatures of their sources. By studying the spectrum of the
background it should be possible to infer the underlying population.
One hopes to take a census of compact objects distributed over astronomical,
and even cosmological distances, by studying these backgrounds.

A cosmological background should have been created in the very early
Universe and later amplified, as a result of parametric amplification,
by its coupling to the background gravitational field 
\cite {grishchuk et al,allen}.  Imprint on such a 
background are the physical conditions that existed in the early Universe
as also the nature of the physical processes that produced the background.
Observing such a background is therefore of fundamental importance as this
is the only way we can ever hope to directly witness the birth of the Universe.
The 2.7 degree K microwave background, which is our firm proof that the
Universe was born from that explosive event, is the relic radiation from that
even but it was in equilibrium with the rest of matter for nearly 300,000
years after big bang and is freely traveling to us only since then. 
The GW background, on the other hand, de-couple from 
the rest of matter soon after the big bang, and is therefore carrying
uncorrupt information about the condition of the Universe at its formation.
In any realistic model of the early Universe the energy density in the 
cosmological background is only a very small fraction 
of the critical density of the Universe, indeed less than one part in
$< 10^{15}.$ As a result it is unlikely that ground-based detectors
will see this relic radiation.  The present design of LISA will 
also not enable us to see this background and one may have to 
fly a special space mission tailored to the observation of this 
relic radiation.  The science output from such an observation is so
fundamental that it is well worth the effort.

Phase transitions in the early Universe, inspired by fundamental
particle physics theories, and cosmological strings and domain walls
are also sources of a stochastic background. These processes are expected
to generate a background which has a different spectrum and strength than
the primordial one. It is expected that some of these
backgrounds may contribute as large as $10^{-9}$  of the closure density
of the Universe. Future ground-based detectors will achieve good enough
sensitivity to measure the background at the level of $10^{-9}$--$10^{-8}$
of the closure density. Such measurements will prove to be a good test
bed for these cosmological models.

\section*{Acknowledgements} I would like to thank Sayan Kar and Naresh
Dadhich for invitation to and hospitality at the International Conference
on Gravitation and Cosmology held at IIT Kharagpur. Without Sayan's 
insistence this review wouldn't have been completed; to that I am 
indebted to him.


\begin{thebibliography}{99}
\bibitem {psr1913} J.H. Taylor, {\it Rev. Mod. Phys.} {\bf 66,} 711 (1994).

\bibitem {Schutz 99} B.F. Schutz, {\it Class. Quantum Grav.} {\bf 16} A131, (1999)
gr-qc/9911034 

\bibitem {Thorne 95} K.S. Thorne, in {\em Proceedings of Snowmass 1994 
Summer Study on Particle and Nuclear Astrophysics and Cosmology,} 
E.W. Kolb and R.D. Peccei, Eds., (World Scientific, Singapore, 1995) 
pp 160-184.

\bibitem {Flanagan} \'E.\'E. Flanagan,
{\it Astrophysical sources of gravitational radiation and prospects 
for their detection,} in Proceedings of 15th International 
Conference on General Relativity and Gravitation (GR15),
eds N. Dadhich and J.V. Narlikar (IUCAA, Pune, India, 1998); 
gr-qc/9804024 

\bibitem {grishchuk et al}
L.P. Grishchuk, V.M. Lipunov, K.A. Postnov, M.E. Prokhorov, 
B.S. Sathyaprakash, {\it Gravitational Wave Astronomy: 
In Anticipation of First Sources to be Detected,} 
{\it Phys. Usp.} (in press) astro-ph/0008481 

\bibitem {lasota.marck} J.-A. Marck and J.-P. Lasota, eds
{\it Relativistic gravitation and gravitational radiation}
(Cambridge Univ. Pr., Cambridge, 1997).

\bibitem {bhawal and iyer} B.R. Iyer and B. Bhawal, 
{\it Black holes, gravitational radiation and the universe,}
(Kluwer Academic Press, 1998).

\bibitem {will.81} C.M. Will, {\it Theory and Experiment in Gravitational 
Physics} (Cambridge Univ. Press, Cambridge, 1981).

\bibitem {schutz.tinto} 
B.F. Schutz, and Tinto, M. (1987) {\it Mon. Not. R. Astron. Soc.}
{\bf 224}, 131.

\bibitem {dt88} S.V. Dhurandhar and M. Tinto, {\it  Mon. Not. R. Astron. Soc.}
 {\bf 234}, 663 (1988).

\bibitem {schutz.1985} B.F. Schutz, in {\it A First Course in General 
Relativity} (Cambridge University Press, Cambridge, 1985).

\bibitem{geo} H. L\"uck {\it et al.}, 
{\it Class. Quantum Grav.} {\bf 14}, 1471 (1997).

\bibitem{ligo} A. Abramovici {\it et al.}, {\it Science} {\bf 256}, 325 (1992).

\bibitem{virgo} B. Caron {\it et al.}, {\it Class. Quantum Grav.}
{\bf 14}, 1461 (1997).

\bibitem{tama}
K. Tsubono, in {\em First Edoardo Amaldi Conference on Gravitational Wave
Experiments,}  (Singapore: World Scientific, 1995) p.\ 112.

\bibitem{lisa}
Bender, P. {\it et al.} {\it {\small LISA}: Pre-Phase A Report,} MPQ 208
(Max-Planck-Institut f\"ur Quantenoptik, Garching, Germany). (Also see
the Second Edition, July 1998.)

\bibitem {binary confusion}  
D. Hils, P.L. Bender, R.F. Webbink, {\it Astrophys. J.} {\bf 360} 75 (1990).

\bibitem {binary radiation reaction}
T. Damour and N. Deruelle, {\it Phys. Lett.} {\bf 87A}, 81 (1981);
and C.R. Acad. Sci. Paris {\bf 293} (II) 537 (1981);
T. Damour, C.R. Acad. Sci.  Paris
{\bf 294} (II) 1355 (1982); and in {\it Gravitational Radiation},
ed. N. Deruelle and T. Piran, pp 59-144 (North-Holland, Amsterdam, 1983).

\bibitem {binary inspiral PN} 
L. Blanchet, T. Damour, B.R. Iyer, C.M. Will and A.G.
Wiseman, {\it Phys. Rev. Lett.} {\bf 74}, 3515 (1995);
L. Blanchet, T. Damour and B.R. Iyer, {\it Phys. Rev.} {\bf D51},
5360 (1995); C.M. Will and A.G. Wiseman, {\it Phys. Rev.} {\bf D54}, 4813 (1996);
L. Blanchet,  B.R. Iyer, C.M. Will and A.G. Wiseman, 
{\it Class. Quantum Grav.} {\bf 13}, 575, (1996); L. Blanchet, 
{\it Phys. Rev.} {\bf D54}, 1417 (1996);

\bibitem{DIS1} T. Damour, B.R. Iyer, B.S. Sathyaprakash, 
{\it Phys. Rev.} {\bf D57}, 885 (1998); ibid {\bf 62}, 084036 (2000);

\bibitem{BD1 and 2}
A. Buonanno and T. Damour, {\it Phys. Rev.} {\bf D59}, 084006 (1999);
ibid, {\bf 62}, 064015 (2000).

\bibitem{DIS3} T. Damour, B.R. Iyer, B.S. Sathyaprakash, 
{\it Phys. Rev. D} (in press); gr-qc/0010009.

\bibitem {will.98} C.M. Will, {\it Phys. Rev.} {\bf D57,} 2061 (1998).

\bibitem {lbbs95} L. Blanchet and B.S. Sathyaprakash,
{\it Class. Quantum Grav.} {\bf 11}, 2807 (1994);
{\it ibid,} {\it  Phys. Rev. Lett.} {\bf 74,} 1067 (1995).

\bibitem {schutz.nature} B.F. Schutz, {\it Nature} {\bf 323,} 310 (1986).

\bibitem {bhevidence} 
M.J. Rees, {\it Class. Quantum Grav.} {\bf 14,} 1411 (1997).

\bibitem {S2:echeverria} F. Echeverria {\it Phys. Rev. } {\bf D40} 3194 (1989).

\bibitem {Flanagan&Hughes98} 
\'E.\'E. Flanagan and S. Hughes, {\it Phys. Rev. } {\bf D57,}
4535 (1998).

\bibitem {cfs} S. Chandrasekhar,
{\em Phys. Rev. Lett.} {\bf 24}  611 (1970);     
J.L. Friedman and B.F. Schutz, 
{\em Astrophys. J.} {\bf 222}  281 (1978).

\bibitem {andersson} N. Andersson,
{\em Astrophys. J.} {\bf 502} 708 (1998);

\bibitem{lindblom et al} L. Lindblom, B.J. Owen and S.M. Morsink 
{\em Phys. Rev. Lett.} {\bf 80} 4843 (1998).

\bibitem {ak review} N. Andersson and K. Kokkotas,
{\em Int. J. Mod. Phys.} (in press); gr-qc/0010102.

\bibitem {allen} B. Allen in Ref.~[6]. 

\end{thebibliography}
\end{document}